# More ferroelectrics discovered by switching spectroscopy piezoresponse force microscopy?

Hongchen Miao[1], Chi Tan[1], Xilong Zhou[1], Xiaoyong Wei[2], Faxin Li[1,3,a]

[1]LTCS and College of Engineering, Peking University, Beijing, 100871, China

[2]Electronic Materials Research Laboratory, Key Laboratory of the Ministry of Education and International Center for Dielectric Research, Xi'an Jiaotong University, Xi'an, 710049, China

[3]HEDPS and Center for Applied Physics and Techniques, Peking University, Beijing, China

**Abstract**

The local hysteresis loop obtained by switching spectroscopy piezoresponse force microscopy (SS-PFM) is usually regarded as a typical signature of ferroelectric switching. However, such hysteresis loops were also observed in a broad variety of non-ferroelectric materials in the past several years, which casts doubts on the viewpoint that the local hysteresis loops in SS-PFM originate from ferroelectricity. Therefore, it is crucial to explore the mechanism of local hysteresis loops obtained in SS-PFM testing. Here we proposed that non-ferroelectric materials can also exhibit amplitude butterfly loops and phase hysteresis loops in SS-PFM testing due to the Maxwell force as long as the material can show macroscopic *D-E* hysteresis loops under cyclic electric field loading, no matter what the inherent physical mechanism is. To verify our viewpoint, both the macroscopic *D-E* and microscopic SS-PFM testing are conducted on a soda-lime glass and a non-ferroelectric dielectric material $Ba_{0.4}Sr_{0.6}TiO_3$. Results show that both materials can exhibit *D-E* hysteresis loops and SS-PFM phase hysteresis loops, which can well support our viewpoint.

---

[a]Author to whom all correspondence should be addressed, Email: lifaxin@pku.edu.cn

In the past decades, piezoresponse force microscopy (PFM) has become a powerful tool to study the electromechanical coupling in piezoelectric and ferroelectric materials at nanoscale[1, 2]. Particularly, the switching spectroscopy PFM (SS-PFM) was widely used to study the microstructure evolution in ferroelectrics, such as domain wall dynamics, nucleation, imprint, etc.[2, 3]. The local phase hysteresis loop in SS-PFM was usually regarded as a typical signature of ferroelectric polarization switching and even used to identify ferroelectricity in biological materials [4-6]. However, recently, such hysteresis loops were also observed in a broad variety of non-ferroelectric materials including doped ZnO[7], $LaAlO_3/SrTiO_3$ heterostructures [8], cellular polypropylene (PP) electrets films[9], lithium-ion battery cathode[10], glass [11, 12] and silicon [13]. These experimental observations cast doubts on the viewpoint that the local hysteresis loops in SS-PFM originate from ferroelectricity. Therefore, it is crucial to explore the general mechanism of local hysteresis loops obtained in SS-PFM testing.

In this letter, we firstly briefly reviewed the principle of SS-PFM testing used in ferroelectrics. Then we proposed that non-ferroelectric materials can also exhibit amplitude butterfly loops and phase hysteresis loops in SS-PFM testing due to the Maxwell force as long as the material can show macroscopic *D-E* hysteresis loops under cyclic electric field loading, no matter what the inherent physical mechanism is. Finally, both the SS-PFM testing and the macroscopic *D-E* measurement were conducted on a soda-lime glass and a non-ferroelectric dielectric material $Ba_{0.4}Sr_{0.6}TiO_3$ to support our viewpoint.

For piezoelectric materials, the first-order harmonic displacement measured by PFM can be expresses as

$$A\cos\phi = d_{33}V_{AC}Q \qquad (1)$$

where *A* and $\phi$ are amplitude and phase of the first-order harmonic displacement respectively. $d_{33}$ and $V_{AC}$ are the piezoelectric coefficient of material and the applied AC voltage, and $Q$ is the quality factor. When the resonance enhancement method is used, the typical $Q$ values for PFM cantilevers in air range from several 10 to 100. For ferroelectric materials, it is well known that $d_{33}$ is proportional to the net polarization $P$, i.e., $d_{33} \propto M_{33}\kappa_3 P$, where $\kappa_3$ and $M_{33}$ are permittivity and electrostrictive coefficient respectively. Thus it is obvious that the curve of $d_{33}$ versus DC electric field in ferroelectrics will be a hysteresis loop due to polarization switching, if the applied field is above the coercive value. Such $d_{33}$ hysteresis loops had been observed in lead zirconate titanate ceramics both at macrosacle [14] and at nanoscale [15, 16]. In SS-PFM testing, a DC electric field with the triangle saw-tooth waveform is applied to the sample and an AC voltage is superposed onto the DC voltage to detect the sample's PFM responses at different DC bias fields, as seen in Fig. 1(a). According to Eq. (1), the local piezoelectric response ($A\cos\phi$) versus DC applied field will be a hysteresis loop for ferroelectrics, as shown in Fig. 1(b). Generally, the curves are measured at the "OFF" state, which can minimize the effects of electrostatic interactions[17]. Obviously, the amplitude is always the absolute value of the deformation while the positive and negative strain will induce 180° phase contrast. Therefore, the local hysteresis loop can be divided into an amplitude ($A$) butterfly loop and a 180° phase ($\phi$) switching hysteresis loop, as shown in Fig. 1(c) and (d).

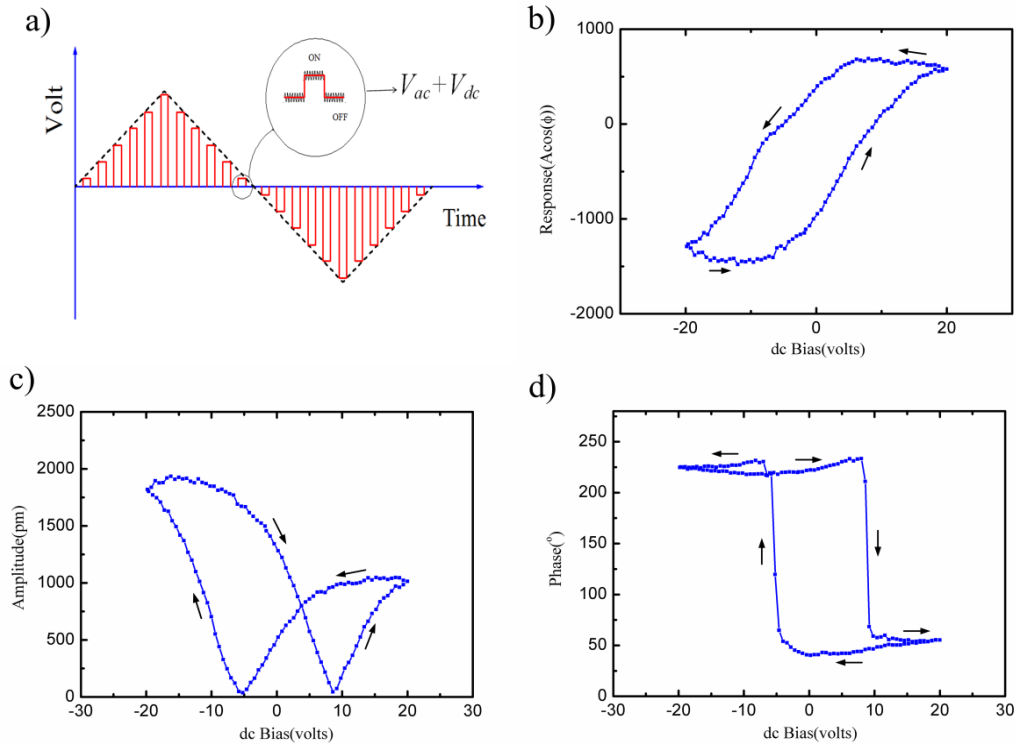

Figure 1: (a) the schematic applied voltage waveform in SS-PFM testing;(b) the response

($A\cos\phi$) hysteresis loop, (c) amplitude butterfly loop and (d) phase switching hysteresis loop of a typical ferroelectric material in SS-PFM testing.

To investigate other originals of SS-PFM hysteresis loops instead of ferroelectric switching, here we consider a non-ferroelectric dielectric material. When a DC voltage $V_{DC}$ is applied uniformly to the dielectric material (which is the case of a dielectric sample with top electrode under SS-PFM testing), surface charge density will be induced on the material which can be expressed by

$$D_{DC} = f(V_{DC}/d) \tag{2}$$

Where $d$ is the thickness of the sample. According to the Coulomb's Law, an equivalent pressure $p_{eq}$ or a Maxwell stress is applied to the sample as follows:

$$p_{eq} = \frac{D_{DC}^2}{\varepsilon_0 \varepsilon_r} \tag{3}$$

Then the sample will shrink along the thickness direction with the strain of

$$S_{DC} = \frac{p_{eq}}{Y} = \frac{D_{DC}^2}{\varepsilon_0 \varepsilon_r Y} \tag{4}$$

where $Y$ is the Young's modulus of the material. When a small AC electric field $V_{AC}$ is superposed onto the DC voltage, additional surface charge will be induced, i.e.,

$$D = D_{DC} + D_{AC} \tag{5}$$

and

$$D_{AC} = \kappa \cdot V_{AC}/d \tag{6}$$

Where $\kappa$ is the dielectric coefficient of the sample under the bias field of $V_{DC}/d$. Then additional strain will be generated in the thickness direction, i.e.,

$$S = \frac{(D_{DC} + D_{AC})^2}{\varepsilon_0 \varepsilon_r Y} \tag{7}$$

and

$$\Delta S = S - S_{DC} \approx \frac{2D_{DC}}{\varepsilon_0 \varepsilon_r Y} \kappa \frac{V_{AC}}{d} \tag{8}$$

Thus, the additional strain is proportional to the applied AC voltage. That is, a non-piezoelectric dielectric material under a bias DC field is equivalent to a piezoelectric material with the equivalent $d_{33}$ of $\frac{2D_{DC}}{\varepsilon_0 \varepsilon_r Y} \kappa$, which is very similar to the apparent piezoelectricity in electrets[18].

Now let us estimate the value of the equivalent $d_{33}$ for typical non-piezoelectric dielectric material. The constant $\kappa$ is close to or on the same order of $\varepsilon_0 \varepsilon_r$, thus $d_{33} \sim 2\frac{D_{DC}}{Y}$. For a low-$\kappa$ dielectric material, the value of $D_{DC}$ even under a high DC field, say $10\,\text{kV/cm}$, is typically less than $0.001\,\text{C/m}^2$. Then, for a moderate-modulus material, say epoxy with the Young's modulus of several $GPa$, the equivalent $d_{33}$ is only on the order of $1\,\text{pC/N}$. Therefore, for hard materials with the Young's modulus of tens of or even $100-200 GPa$, the equivalent $d_{33}$ is even smaller and negligible. While for soft dielectrics, such as the cellular polypropylene (PP) electrets films, the Young's modulus is on the order of $\sim MPa$, which makes the equivalent $d_{33}$ reach up to several $100\,\text{pC/N}$ [19].

However, the case in SS-PFM testing is somewhat different. Typically, the testing sample is top-electrode free and the electric field generated by the AFM tip is highly concentrated which can reach up to $10^{10}\,\text{V/m}$ beneath the tip with several tens of DC voltages on[20]. Furthermore, the electric field will attenuate by two orders of magnitude within $\leq 1\mu m^3$ inside the material[21]. In this case, the surface charge density beneath the tip will saturate and may reach $0.1 \sim 1.0\,\text{C/m}^2$ in the small area. The additional electric field induced by the AC voltage may reach about $10^9\,\text{V/m}$. On the other hand, because the Maxwell stress only apply at the small area, the clamping effect from the neighboring material makes the real deformation drastically reduced by three orders [22] in the case of linear elastic deformation. Here we use the attenuation factor $K$ to describe the clamping effect. Fortunately the resonance technique in PFM can enhance the vibration signal by a factor $\sim Q$, whose value typically ranges from several 10 to 100. Taking into all the above factors, the response detected by PFM can be expressed by

$$A\cos\phi = \Delta S \cdot d_1 \approx KQd_1 \frac{2D_{DC}}{Y} E_{AC} \tag{9}$$

Where $d_1$ is the thickness of thin layer beneath the surface in which the electric field is highly concentrated, $E_{AC}$ is the average of concentrated AC electric field inside this thin layer. Taking the typical values of $K=0.001$, $Q=100$, $d=1\,\mu m$, $D_{DC}=0.5\,C/m^2$, $Y=100\,GPa$ (hard material), $E_{AC}=10^9\,V/m$, the vibration amplitude estimated by Eq.(9) can reach $\sim nm$, which is large enough to be detected by PFM. Therefore, we can deduce from Eq.(9) that non-piezoelectric hard dielectrics can also show significant PFM responses at the "ON" state in SS-PFM testing if the resonance enhancing technique is used. Furthermore, it can be seen that if the sign of the surface charge reverses, the PFM phase will change by 180°.

For those dielectric materials without *D-E* hysteresis, the surface charge will vanish immediately after removing the applied field, thus the sample will show little PFM responses at the "OFF" state in SS-PFM testing. Furthermore, it can also be deduced from Eq.(9) that when the applied voltage reverses, the PFM phase will undergo a sharp change by 180° at 0 V.

However, for the dielectrics with obvious *D-E* hysteresis loops, as shown in Fig.2 (a), remnant surface charge still exists upon removing the DC applied field, which makes the material show significant PFM responses at the "OFF" state in SS-PFM testing. When the applied DC voltage reverses and reaches a certain value, say Point C in Fig.2 (b), the surface charge vanishes before its sign changes, leading to the little PFM amplitude and the PFM phase reversal at this point. Then, during a cycle of the triangle saw-tooth waveform DC field in SS-PFM testing, phase hysteresis loops and amplitude butterfly loops will appear in these kinds of dielectrics.

Bearing in mind that in above analysis no specific mechanism is presumed for the macroscopic *D-E* hysteresis loops in dielectrics, here we propose a general rule for the observed SS-PFM local hysteresis loops in dielectrics. That is, a dielectric material can exhibit local phase hysteresis loops and amplitude butterfly loops in SS-PFM testing as long as the material can show macroscopic *D-E* hysteresis loops under cyclic electric field loading, no matter what the inherent physical mechanism is. Note that this rule also applies for ferroelectrics with polarization switching. In ferroelectrics, polarization switching occurs at the coercive field which will change the sign of the surface charge, leading to the PFM phase change by 180°. The corresponding analysis is similar and will not be iterated here.

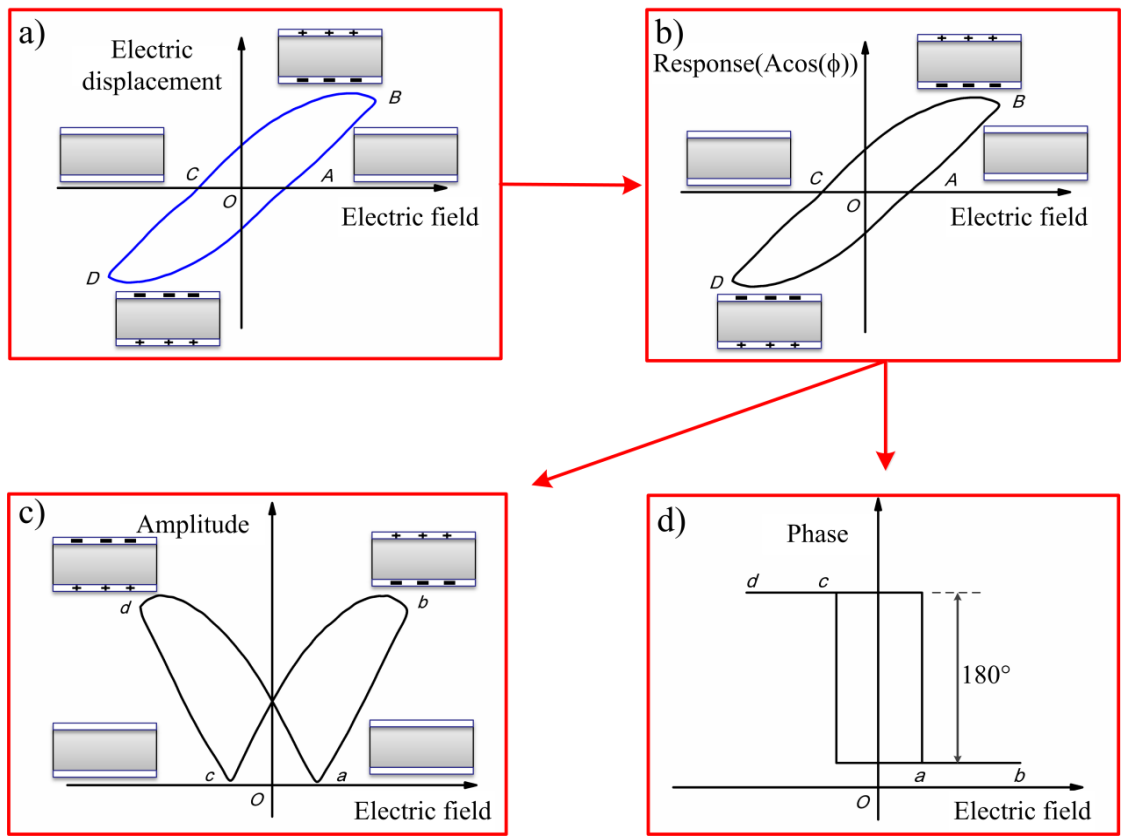

Figure 2: Illustration of the mechanism of local hysteresis loops in non-ferroelectric materials obtained by SS-PFM testing. (a) macroscopic *D-E* hysteresis loop; (b) response hysteresis loop; (c) amplitude butterfly loop and (d) phase hysteresis loop obtained in SS-PFM testing.

It should be noted that the in the proposed rule, the *D-E* hysteresis loops should be caused by the intrinsic properties of the materials, such as dielectric relaxation or other inherent mechanism caused surface charge lag. The charge leakage induced apparent *D-E* hysteresis loops measured using conventional Sawyer-Tower circuit should excluded because in this case the measured electric displacement (D) is mostly contributed by the integral of the leakage current, not the surface charge. The charge leakage induced *D-E* hysteresis loops can be distinguished from the intrinsic *D-E* hysteresis based on the loop shape, which had been clearly indicated by Scott[23].

To support our viewpoint proposed above, we firstly conduct both the macroscopic *D-E* hysteresis measurement and SS-PFM testing on a 1mm-thick dielectric soda-lime glass. In this work, to remove the effect of charge leakage on the macroscopic *D-E* hysteresis loops, the similar waveform electric field with that in SS-PFM testing was employed in the measurements which were realized by using a Radiant RC ferroelectric analyzer with the maximum applied voltage of $10\,\text{kV}$. The SS-PFM testing were conducted based on a commercial AFM (Asylum Research MFP-3D) using conductive probes (Olympus AC240) with the nominal spring constant of $2\,\text{N/m}$ and the first free resonance of $\sim 70\,\text{kHz}$. The maximum applied DC bias voltage is $200\,\text{V}$ and the detecting AC voltage of $3\,\text{V}$.

Fig. 3(a) shows that the soda-lime glass exhibits macroscopic *D-E* hysteretic loops even at the "OFF" state, which indicates that the surface charge hysteretic response to the applied field in soda-lime glass is an intrinsic material property. Thus it is straightforward to infer that the surface charges response to the DC electric field will also be hysteretic in the SS-PFM testing. Then according to Eq. (9), the soda-lime glass should exhibit the local hysteresis loop in SS-PFM testing and this had been confirmed by the measured hysteresis curves in Fig. 3(b), (c), (d), which are nearly identical to those obtained on a typical ferroelectric material.

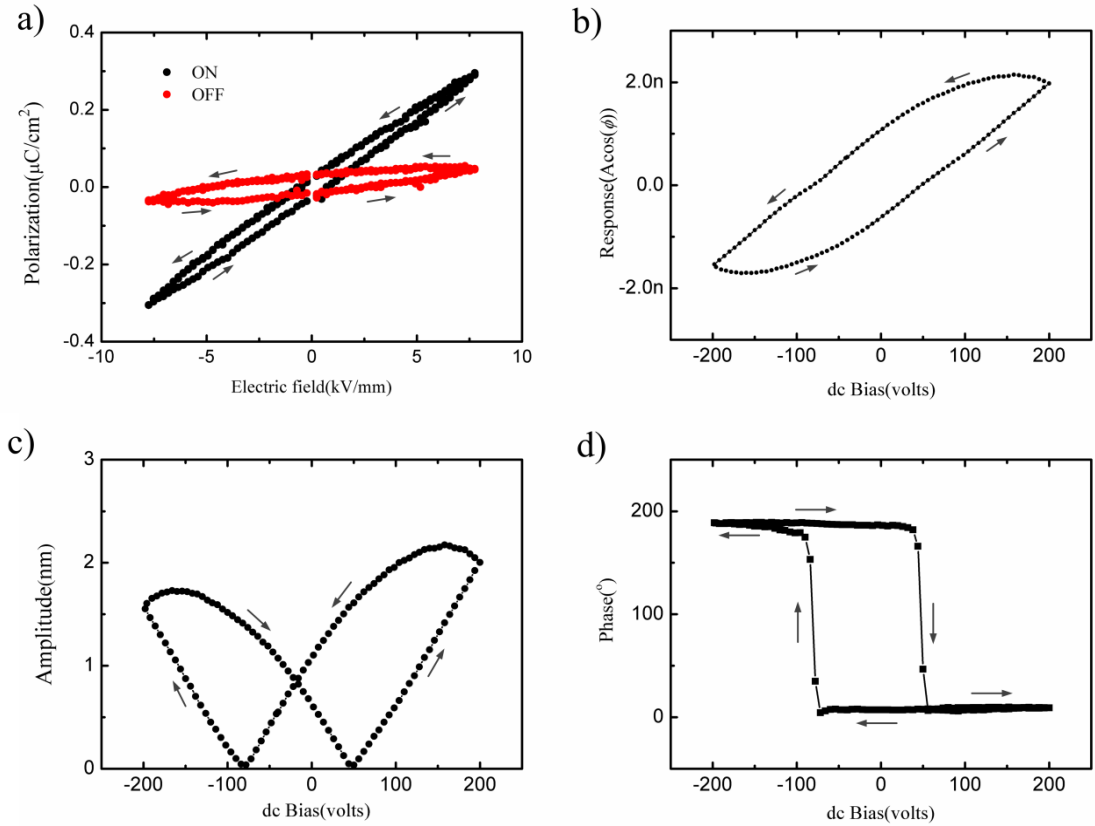

Figure 3: Hysteresis loops of a soda-lime glass. (a) macroscopic $D$–$E$ hysteresis loop measured using the similar waveform field with that in SS-PFM testing; (b) the local hysteresis loop, (c) amplitude butterfly loop and (d) phase switching hysteresis loop measured in SS-PFM testing.

To further examine our viewpoint, we choose another purely dielectric material, barium strontium titanate, $Ba_{1-x}Sr_xTiO_3$ (BST), which is the solid solution phase between $BaTiO_3$ and $SrTiO_3$. BST with $x \geq 0.4$ is cubic symmetry at room temperature and its curie temperature is about 210K [24], therefore, $Ba_{0.4}Sr_{0.6}TiO_3$ ceramic is a non-ferroelectric dielectric material [25]. The $D$-$E$ hysteresis loops are found at the "OFF" state in the $Ba_{0.4}Sr_{0.6}TiO_3$ ceramic, as shown in Fig. 4(a). As expected, it exhibits the local hysteresis loop in SS-PFM experiment, as shown in Fig. 4(b), 4(c) and 4(d), which can well support our viewpoint.

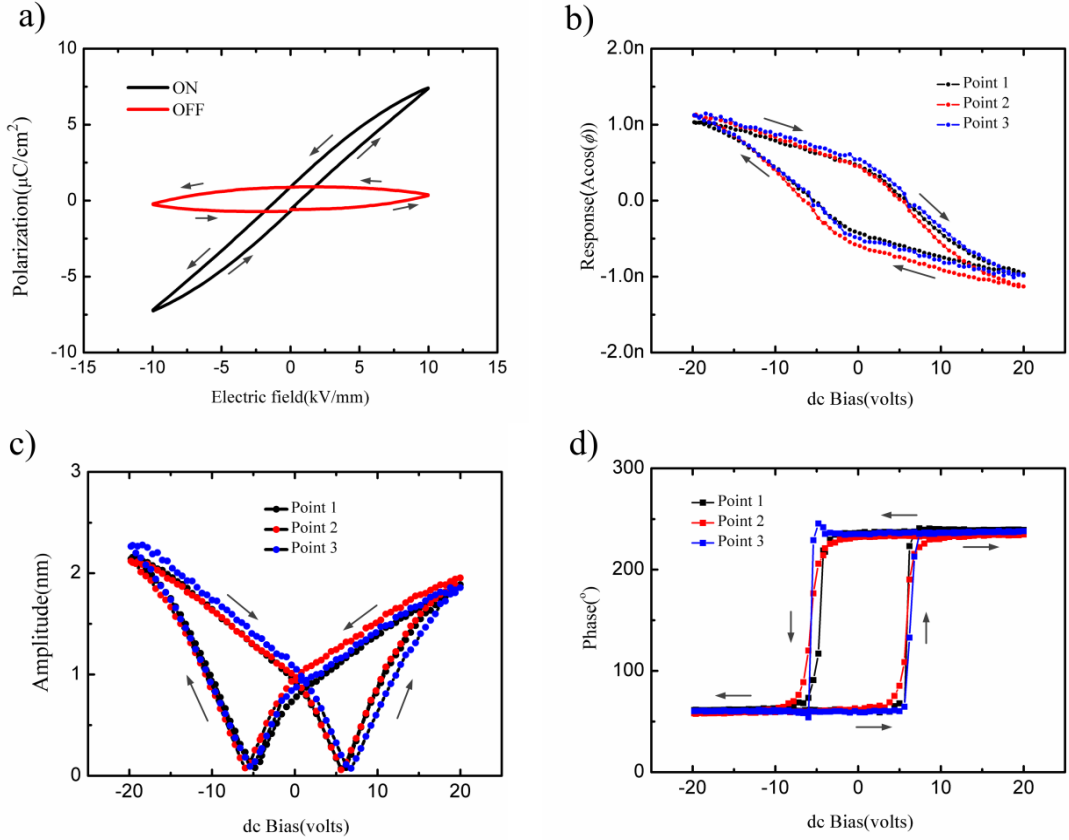

Figure 4: Hysteresis loops of a non-ferroelectric dielectric $Ba_{0.4}Sr_{0.6}TiO_3$ ceramic. (a) Macroscopic $D-E$ hysteresis loop measured using the similar waveform field with that in SS-PFM; (b) the local hysteresis loop, (c) amplitude butterfly loop and (d) phase switching hysteresis loop measured in SS-PFM testing.

The proposed rule is also valid for those materials with polarization switching, including ferroelectrics and ferroelectrets, in which both the macroscopic $D-E$ hysteresis loops and the SS-PFM local hysteresis loops had been observed elsewhere[9]. The rule should also apply for the material characteristic of resistive switching, such as doped ZnO[7], whose dielectric constant switches with applied voltage and it is expected to exhibit macroscopic $D-E$ hysteresis loops. For the electrochemical reaction materials, the ion migration will also induced surface charge variations and typically the ion migration lags the applied voltage, which makes the materials should also show macroscopic $D-E$ hysteresis loops and SS-PFM loops[10]. The Vegard effect[26] will enhance the PFM responses, but it will not induce the SS-PFM loops itself, just like the piezoelectric effect in ferroelectrics.

As to those biomaterials which exhibit local SS-PFM hysteresis loops[4-6], they are porous materials with space charges and typically treated as bio-electrets, very similar to the ferroelectrets and should also show ferroelectric-like *D-E* hysteresis loops. That is, the proposed rule should also be applicable for those biomaterials with reversible space charge. However, in most biomaterials, charge leakage is very severe and the intrinsic *D-E* curves are rather difficult to measure even using the waveform field similar with that in SS-PFM testing.

In summary, we proposed a general rule that non-ferroelectric materials can also exhibit amplitude butterfly loops and phase switching loops in SS-PFM testing due to the Maxwell force as long as the material can show macroscopic *D-E* hysteresis loops under cyclic electric field loading. Experimental results of non-ferroelectric glasses and a dielectric material $Ba_{0.4}Sr_{0.6}TiO_3$ can well support our viewpoint. The proposed rule indicates that SS-PFM testing cannot be used to identify ferroelectricity as the local hysteresis loop is not the unique feature of ferroelectric polarization switching, just like the macroscopic *D-E* hysteresis measurement method as pointed out by Scott[23]. The viewpoint proposed in this letter should be helpful to the scholars in PFM and ferroelectrics field.

**Acknowledgement**

FL gratefully thanks Professor James F. Scott (Cambridge University) for the helpful discussions during his short visit to Peking University in June 2014.

**References**
[1] Bonnell D. A., Basov D. N., Bode M., Diebold U., Kalinin S. V., Madhavan V., Novotny L., Salmeron M., Schwarz U. D. and Weiss P. S., *Rev. Mod. Phys.,* **84** (2012) 1343.
[2] Balke N., Bdikin I., Kalinin S. V. and Kholkin A. L., *J. Am. Ceram. Soc.,* **92** (2009) 1629-47.
[3] Soergel E., *J. Phys. D. Appl. Phys.,* **44** (2011) 464003.
[4] Liu Y. M., Zhang Y. H., Chow M. J., Chen Q. N. and Li J. Y., *Phys. Rev. Lett.,* **108** (2012) 078103.
[5] Zhou X. L., Miao H. C. and Li F. X., *Nanoscale,* **5** (2013) 11885-93.
[6] Heredia A., Meunier V., Bdikin I. K., Gracio J., Balke N., Jesse S., Tselev A., Agarwal P. K., Sumpter B. G., Kalinin S. V. and Kholkin A. L., *Adv. Funct. Mater.,* **22** (2012) 2996-3003.
[7] Herng T. S., Kumar A., Ong C. S., Feng Y. P., Lu Y. H., Zeng K. Y. and Ding J., *Scientific Reports,* **2** (2012) 1038.
[8] Bark C. W., Sharma P., Wang Y., Baek S. H., Lee S., Ryu S., Folkman C. M., Paudel T. R., Kumar

A., Kalinin S. V., Sokolov A., Tsymbal E. Y., Rzchowski M. S., Gruverman A. and Eom C. B., *Nano Letters,* **12** (2012) 1765-71.

[9] Miao H. C., Sun Y., Zhou X. L., Li Y. W. and Li F. X., *J. Appl. Phys.,* **116** (2014) 066820.

[10] Balke N., Jesse S., Morozovska A. N., Eliseev E., Chung D. W., Kim Y., Adamczyk L., Garcia R. E., Dudney N. and Kalinin S. V., *Nature Nanotechnology,* **5** (2010) 749-54.

[11] Proksch R., *arXiv,* **1312.6933**

[12] Chen Q. N., Ou Y., Ma F. Y. and Li J. Y., *Appl. Phys. Lett.,* **104** (2014) 242907.

[13] Sekhon J. S., Aggarwal L. and Sheet G., *Appl. Phys. Lett.,* **104** (2014) 162908.

[14] Zhang Y., Baturin I. S., Aulbach E., Lupascu D. C., Kholkin A. L., Shur V. Y. and Rodel J., *Appl. Phys. Lett.,* **86** (2005) 012910.

[15] Hidaka T., Maruyama T., Saitoh M., Mikoshiba N., Shimizu M., Shiosaki T., Wills L. A., Hiskes R., Dicarolis S. A. and Amano J., *Appl. Phys. Lett.,* **68** (1996) 2358-9.

[16] Bdikin I. K., Kholkin A. L., Morozovska A. N., Svechnikov S. V., Kim S. H. and Kalinin S. V., *Appl. Phys. Lett.,* **92** (2008) 182909.

[17] Jesse S., Lee H. N. and Kalinin S. V., *Rev. Sci. Instrum.,* **77** (2006).

[18] Hillenbrand J. and Sessler G. M., *Ieee. T. Dielect. El. In,* **7** (2000) 537-42.

[19] Wegener M. and Bauer S., *Chemphyschem,* **6** (2005) 1014-25.

[20] Tian L. L., Aravind V. R. and Gopalan V. Quantitative Piezoresponse Fore Microscopy: Calibrated Experiments,Analytical Theory and Finite Element Modeling. In: Kalinin S. V., Gannepalli A., editors. Scanning Probe Microscopy of Functional Materials. London and New York: Springer; 2010.

[21] Otto T., Grafstrom S. and Eng L. M., *Ferroelectrics,* **303** (2004) 747-51.

[22] Jungk T., Hoffmann A. and Soergel E., *Appl. Phys. A-mater.,* **86** (2007) 353-5.

[23] Scott J. F., *J. Phys-condens. Mat.,* **20** (2008) 021001.

[24] Kim S. W., Choi H. I., Lee M. H., Park J. S., Kim D. J., Do D., Kim M. H., Song T. K. and Kim W. J., *Ceram. Int.,* **39** (2013) S487-S90.

[25] Zhou L. Q., Vilarinho P. M. and Baptista J. L., *J. Eur. Ceram. Soc.,* **19** (1999) 2015-20.

[26] Kalinin S., Balke N., Jesse S., Tselev A., Kumar A., Arruda T. M., Guo S. L. and Proksch R., *Mater. Today* **14** (2011) 548.